\documentclass[aps,preprintnumbers,superscriptaddress,twocolumn,amsmath,amssymb,floatfix,prb]{revtex4} 
\pdfoutput=1
\usepackage{graphicx}
\usepackage[usenames,dvipsnames]{color}
\usepackage[colorlinks,bookmarks=false,citecolor=NavyBlue,linkcolor=Red,urlcolor=blue]{hyperref}
\newcommand\be{\begin{equation}}
\newcommand\bea{\begin{eqnarray}}
\newcommand\bes{\begin{subequations}}
\newcommand\esu{\end{subequations}}
\newcommand\ee{\end{equation}}
\newcommand\eea{\end{eqnarray}}

\def\doi{http://dx.doi.org/}

\newcommand\ocite {\onlinecite}

\begin{document}

\title{Quenches from bosonic gaussian initial states to the Tonks-Girardeau limit:\\
stationary states and effects of a confining potential }

\author{Alvise Bastianello}
\affiliation{SISSA \& INFN, via Bonomea 265, 34136 Trieste, Italy}

\author{Mario Collura}
\affiliation{The Rudolf Peierls Centre for Theoretical Physics, 
Oxford University, Oxford, OX1 3NP, United Kingdom.}

\author{Spyros Sotiriadis}
\affiliation{Institut de Math\'ematique de Marseille, (I2M) Aix Marseille Universit\'e, CNRS,
Centrale Marseille, UMR 7373, 39 rue F. Joliot Curie, 13453, Marseille, France}
\affiliation{University of Roma Tre, Department of Mathematics and Physics, 
L.go S. L. Murialdo 1, 00146 Roma, Italy}


\begin{abstract}
We consider the general problem of quenching an interacting Bose gas from 
the noninteracting regime to the strongly repulsive limit described by the Tonks-Girardeau gas, 
with the initial state being a gaussian ensemble in terms of the bosons. 
A generic multi-point correlation function in the steady state can be fully described in terms 
of a Fredholm-like determinant suitable both for a numerical and an analytical study in certain limiting cases. 
Finally, we extend the study to the presence of a smooth confining potential 
showing that, in the thermodynamic limit, the time evolution of the two-point function 
can be mapped to a classical problem.

\end{abstract}


\maketitle
\section{Introduction}
The recent advances in experimental techniques concerning the realm of cold atomic gases \cite{exp} made it urgent to reach a better understanding of the behavior of closed quantum systems when driven out of equilibrium. 
A paradigmatic out-of-equilibrium protocol consists in the so called \emph{quantum quench}\cite{calabrese-cardy},  i.e. the system starts in the ground state, an excited state or a thermal state of its Hamiltonian 
and then some parameter of the Hamiltonian is abruptly changed, resulting in non-trivial dynamics. 
Despite the fact that the whole system is closed and so the time evolution is non dissipative, finite subsystems are expected to relax towards a steady state and immense efforts have been devoted to investigating this hypothesis\cite{pssv}.

Particularly appealing is the world of one dimensional systems, due to the existence of \emph{integrable models}\cite{Korepin}, i.e. those that possess infinitely many local integrals of motion in addition to the Hamiltonian which allow an exact solution of the dynamics\cite{integrability_review}. 
The presence of infinitely many conserved quantities makes the steady state retain additional information of the initial state compared to the non-integrable case so that the Gibbs Ensemble needs to be modified to a Generalized Gibbs Ensemble (GGE) density matrix\cite{gge}
\be
\rho_{\text{GGE}}=\frac{1}{\mathcal{Z}}e^{-\sum_j \lambda_j \mathcal{I}_j}\, ,
\ee
where $\mathcal{I}_j$ are the relevant conserved charges and $\lambda_j$ the associated Lagrange multipliers.
The correctness of the GGE prediction has been verified in many different cases\cite{gge_work} and recent works focused in understanding the locality features of the charges to be included in the GGE construction\cite{localcharges}.

Despite several achievements in the study of integrable models out-of-equilibrium, apart from free models analytic results have been derived only  for restricted classes of observables and initial states and at the price of tremendous efforts\cite{quench_initial_states}. Recent investigations probed inhomogeneous quench protocols, obtaining analytical results in the framework of transport problems\cite{transport}.

In between the complexity of quenching a truly interacting integrable model and the simplicity of a free model lies a special interaction quench in the well known Lieb Liniger model\cite{Korepin,quench_liebliniger}.
Its Hamiltonian describes a one-dimensional gas of non-relativistic bosons with mass $m$ interacting through a density-density contact potential of strength $c$
\be
H_\text{LL} = \int dx \left[ \frac{  \partial_x \psi^{\dag}(x) \partial_x \psi(x)}{2m} 
+ c \, \psi^{\dag}(x) \psi^{\dag}(x) \psi(x) \psi(x) \right]\, .\label{LL}
\ee
Above, the scalar fields $\psi^{\dag}(x)$ satisfy the canonical bosonic commutation
relations $[\psi(x),\psi^{\dag}(y)] = \delta(x-y)$.
For arbitrary values of $c$ the model is a truly interacting integrable model, however in two special limits the dynamics is that of free particles: at $c=0$ the model reduces to free bosons, while in the $c\to\infty$ limit known as Tonks-Girardeau gas (TG) the bosons behave as \emph{free} but impenetrable particles \cite{tonksgas} that are equivalent to \emph{free fermions} by means of a Jordan Wigner transformation. Compared with many other quenches among free theories, the modes at $c=0$ and $c=\infty$ are not connected through a simple linear relation.

This quench protocol has already been successfully addressed in the literature, 
choosing as initial state the BEC state and analytical expressions for the steady state correlators were obtained\cite{kcc14}. 
Subsequently, these results were generalized to include a hard wall trap\cite{mckc14}.
The presence of simple analytical results for an experimentally relevant 
quench protocol is certainly appealing, 
however the above mentioned works concern a rather peculiar initial state.
Among the flaws of the BEC as pre-quench state are:
{\it(i)} the total absence of any space dependence in the correlation functions 
(except when a confining potential is present); 
{\it(ii)} the strong dependence on the trap shape\cite{mckc14}
which makes the proper thermodynamic limit difficult to define. 
In this sense, the BEC state is rather pathological and therefore, 
it is worth to investigate a different class of states which hold a more realistic behavior. On the other hand, the initial state is still required to be simple enough to allow an analytical treatment.
A good compromise that meets both requirements is given by the {\it gaussian} (in terms of the bosons) ensembles, which are the initial states we consider in this paper.
In particular, we present a detailed analysis of the multipoint correlation functions in the steady state attained at late times, with emphasis on the large distance decaying properties. 
We initially consider an homogeneous system and subsequently a thermodynamically large trap, providing closed expressions for the correlators suitable both for a numerical and analytical study.
The correlation functions are notoriously difficult to be obtained in quenches in truly interacting integrable models: the simplification of quenching towards the Tonks-Girardeau limit makes these quantities more easily accessible.
While our results apply to arbitrary gaussian (in terms of the bosons) initial states, we consider the prototypical example of the \emph{free thermal state}, i.e. an ensemble with Bose-Einstein distribution: the density-density correlation function in the steady state is proven to decay exponentially, the decay length being a non trivial function of the temperature and chemical potential.
When the zero temperature limit is taken, the decay behavior changes from exponential to algebraic: actually, in the zero temperature limit the \emph{exact} correlation function is computable. This result is striking different from what is obtained quenching from a BEC \cite{kcc14}, where an exponential decay is observed, but this is attributed to the fact that the finite temperature Gibbs Ensemble is not continuously connected to the BEC state as long as one dimension is concerned.
Our discussion is organized as follows.
In Section \ref{Qsetup} we present the general technology needed to handle this quench protocol: under the condition of initial clustering of fermionic correlators, dephasing among the free fermionic modes of the post quench dynamics leads to a \emph{gaussian} steady state\cite{gaussification} in terms of the fermions (i.e. the charges appearing in the GGE are quadratic in terms of the fermionic operators).
Because of the free dynamics connecting the initial state with the steady state, the open technical problems reduce to \emph{(i)} computing the initial fermionic correlators on a bosonic gaussian ensemble and \emph{(ii)} computing back the bosonic correlators on the fermionic gaussian steady state. We derive closed expressions in terms of Fredholm determinants to fulfill this task. 
These results are employed in Section \ref{sec4} to study both numerically and analytically quenches in homogeneous systems, i.e. the particles are confined in a thermodynamically large interval with periodic boundary conditions. 
While being an important benchmark, homogeneous systems with periodic boundary conditions are a crude approximation of realistic experimental systems, in which an inhomogeneous confining potential is always present: 
Section \ref{sec5} extends our study to the presence of a trap. We show how in the thermodynamic limit (which is necessary for the system to exhibit relaxation) a semiclassical time evolution naturally emerges and allows us to consider traps of arbitrary shapes. Our final conclusions are gathered in Section \ref{conclusion}, while Appendix \ref{HBsec}-\ref{WKBsec} contain some technical details in support to the main text.

\section{Quench set up}\label{Qsetup}

In the infinite repulsive regime, the Lieb Liniger gas (\ref{LL}) can be described as a free model of hard-core bosons \cite{hardcore} $\Psi(x)$
\be
H_\text{LL}=\int dx \frac{\partial_x \Psi^\dagger(x) \partial_x \Psi(x)}{2m}, \hspace{2pc} c=+\infty\, .
\ee
The infinite repulsive term prevents the possibility of having two particles in the same position and this can be achieved through a definition of the hard-core bosonic fields $\Psi(x)$, obtained projecting the standard bosonic fields $\psi(x)$ on the subspace with at most one particle in position $x$ 
\be
\Psi(x)=P_x\psi(x)P_x\, .
\ee
Above, $P_x=|0_x\rangle\langle 0_x|+|1_x\rangle \langle 1_x|$ is the projector on the subspace with no more than a particle in position $x$.
The constraint is evident from the commutation rules satisfied by the hard-core bosonic fields, i.e.
\be
\{\Psi(x),\Psi(x)\}=0,\hspace{2pc} \{\Psi(x),\Psi^\dagger(x)\}=1
\ee
at the same position and the standard commutation rules $[\Psi(x),\Psi(y)]=[\Psi(x),\Psi^\dagger(y)]=0$ whenever $x\ne y$ \cite{kcc14}.
Actually, a correct definition of the hard-core bosons requires a proper lattice regularization (see Appendix \ref{HBsec} for further details): in order to consider directly the continuum limit it is convenient to introduce the fermionic field $\Phi(x)$ through a Jordan-Wigner transformation $
\Phi(x) = \exp\left( i\pi \int^{x}dz\, \Psi^{\dag}(z)\Psi(z)\right) \Psi(x)
$. While the hard-core bosons need a lattice regularization, on the fermionic fields the continuum limit can be taken and standard commutation rules $\{\Phi(x),\Phi^{\dag}(y)\} = \delta(x-y)$ are recovered. The strong-interacting Hamiltonian becoming thus
quadratic in terms of those fermionic fields. 

Both models, namely the non-interacting and the strong-interacting, 
are therefore related via a non-local transformation which 
nevertheless makes a well-established connection. 
Whenever we need to pass from one theory to the other,
e.g. via a quench protocol, the problem reduces to compute 
the multipoint correlators of the post-quench fields in terms of the 
pre-quench ones.
This dictionary is provided by the underling regularized hard-core bosons, whose normal ordered correlation functions can be computed as if the hard core bosons were canonical bosonic operators: the details are reported in Appendix \ref{HBsec}, thereafter we limit ourselves to quote the necessary results.
For example, the fermionic multipoint correlation function
can be expressed in terms of bosonic fields as
\bea\label{eq_multipoint_mapping}
&&\langle \Phi^{\dag}(y_1)\dots \Phi^{\dag}(y_l)\Phi(x_l)\dots\Phi(x_1) \rangle = \\
&&\langle : \psi^{\dag}(y_1)\dots \psi^{\dag}(y_l) e^{-2\int_{\mathcal{D}}dz \, n(z)} \psi(x_l)\dots\psi(x_1) : \rangle \nonumber,
\eea
where, without loss of generality, we assumed $y_1 < \dots < y_l$ and $x_1 < \dots < x_l$, 
$n(z) \equiv \psi^{\dag}(z)\psi(z)$ being the density operator, 
and the integration domain $\mathcal{D}$ is defined via
\be
\int_{\mathcal{D}} dz \, n(z) = \int_{-\infty}^{\infty} dz \, n(z) (1-e^{i\pi \sum_{j=1}^{l} \chi_{j}(z) })/2\, ,
\ee
where $\chi_{j}(z)$ is the characteristic function of the interval $(y_{j},x_{j})$.
The same expression can be used to compute the bosonic correlators starting from the fermionic ones, provided in Eq. (\ref{eq_multipoint_mapping}) the bosons are replaced with fermions and vice versa. A rigorous derivation of this relation requires a lattice regularization similarly to what has been done 
in Ref. \ocite{kcc14} and we leave these considerations to Appendix \ref{HBsec}. 

Considering a quench from homogeneous free bosons towards the Tonks-Girardeau limit, the fermions evolve as free fields, therefore the system typically exhibits \emph{gaussification}\cite{gaussification}, i.e. all the fermionic multipoint connected correlators vanish apart from the two point correlator. Assuming translation invariance, the two point correlator is conserved by the time evolution and it is the only information of the initial state that survives in the steady state.

Therefore, the quench problem is reduced to computing the expectation value of the two point correlator on the pre-quench state through Eq. (\ref{eq_multipoint_mapping}):
handling this expression for arbitrary pre-quench states appears as an hopeless task, since even in the computation of the fermionic two point correlator the whole set of multipoint bosonic correlators is required, as it is clear from the power expansion of the exponential in Eq. (\ref{eq_multipoint_mapping}). 
Nevertheless, some special classes of states allow for a brute force computation of Eq. (\ref{eq_multipoint_mapping}), i.e. the already studied BEC case \cite{kcc14} and the gaussian ensembles that we are going to analyze.

Notice that, being the steady state gaussian in terms of the fermionic fields, the bosonic correlators can be obtained solving the inverse problem with respect to that we analyzed so far. Thus, we discuss the computation of fermionic correlators from a gaussian bosonic ensemble: the calculations in the other direction are completely analogous.
Evaluating Eq. (\ref{eq_multipoint_mapping}) on a gaussian ensemble amounts to an extensive use of the Wick theorem, however the calculations are greatly simplified replacing Eq. (\ref{eq_multipoint_mapping}) with an expression in terms of the \emph{connected correlators}
\bea
&&\langle\Phi^\dagger(y)\Phi(x) \rangle=\langle:\psi^\dagger(y)e^{-2\int_\mathcal{D} dz\; n(z)}\psi(x): \rangle= \label{3} \\
&&\langle:\psi^\dagger(y)e^{-2\int_\mathcal{D} dz\; n(z)}\psi(x): \rangle_\text{c}\langle :e^{-2\int_\mathcal{D} dz\; n(z)}:\rangle,\nonumber
\eea
where $\langle...\rangle_\text{c}$ is the connected expectation value. It must be stressed that the connected part is constructed considering the particle density $n(z)$ as a unique operator and not as a composite one $n(z)=\psi^\dagger(z)\psi(z)$. 
The derivation of (\ref{3}) is a simple exercise best understood through an analogy with standard Feynman diagrams where $e^{-2\int_\mathcal{D} dz\; n(z)}$ plays the role of the ``action". In the computation of correlation functions the expectation value of the action factorizes and we are left with the sum of connected diagrams, i.e. (\ref{3}). Besides, $\langle:e^{-2\int_\mathcal{D} dz\; n(z)}:\rangle$ can written in terms of connected correlators thanks to the cumulant expansion \cite{weinberg}
\be
\langle:e^{-2\int_\mathcal{D} dz\; n(z)}:\rangle=e^{\sum_{j=1}^\infty \frac{(-2)^j}{j!}\int_\mathcal{D} d^j z\, \langle :n(z_1)...n(z_j):\rangle_{c}}\, .
\ee

\begin{figure}[t!]
\includegraphics[width=0.45\textwidth]{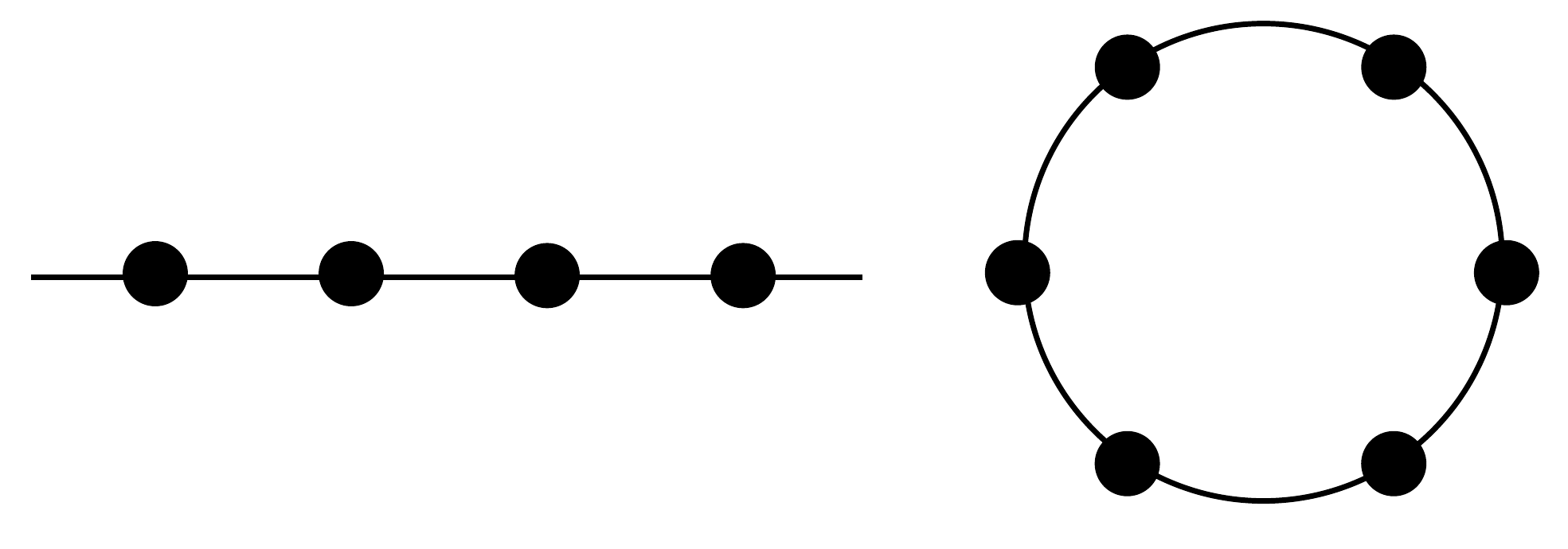}
\caption{\label{figfey} Feynman diagram representation of Eq. (\ref{5}-\ref{6}): black dots represent the density operators $n(z)$. The Wick theorem contracts the fields pairwise: since the density contains two fields $n(z)=\psi^\dagger(z)\psi(z)$ there are two departing legs from each black dot. In Eq. (\ref{5}) external legs are present in association with the fields $\psi^\dagger(y)$ and $\psi(x)$ (Left), while in the diagrams of Eq. (\ref{6}) external legs are absent (Right).}
\end{figure}

The computation of the connected correlators on a gaussian ensemble is straightforward (see Fig. \ref{figfey}) and we reach the following expressions (where $\mathcal{C}(y,x)=\langle\psi^\dagger(y)\psi(x)\rangle$)
\bea
&&\langle:\psi^\dagger(y)e^{-2\int_\mathcal{D} dz\; n(z)}\psi(x): \rangle_\text{c}= \label{5}\\
&&\nonumber\sum_{j=0}^\infty (-2)^j\int_\mathcal{D} d^jz \prod_{l=0}^j\mathcal{C}(z_{l+1},z_l)
\eea
and
\be
\langle :e^{-2\int_\mathcal{D} dz\; n(z)}:\rangle=e^{\sum_{j=1}^\infty \frac{(-2)^j}{j}\int_\mathcal{D} d^jz \prod_{l=1}^j\mathcal{C}(z_{l+1},z_l)}\,\,\, ,\label{6}
\ee
where in the first expression $z_{j+1}=y$ and $z_0=x$, while in the second expression $z_{j+1}=z_1$. 
Eq. (\ref{5}-\ref{6}) are readily resummed in compact expressions once we define $\mathcal{C}_{\mathcal{D}}$ as the linear operator obtained restricting the correlator $\mathcal{C}$ on the domain $\mathcal{D}$, i.e. $\mathcal{C}_\mathcal{D}(s,t)=\mathcal{C}(s,t)$ and $s,t\in \mathcal{D}$.
\be
\langle:\psi^\dagger(y)e^{-2\int_\mathcal{D} dz\; n(z)}\psi(x): \rangle_\text{c}=\left(\frac{\mathcal{C}_\mathcal{D}}{1+2\mathcal{C}_\mathcal{D}}\right)(y,x)\,  \, ,
\ee
\be
\langle:e^{-2\int_\mathcal{D} dz\; n(z)}: \rangle_\text{c}=\det_\mathcal{D} (1+2\mathcal{C}_\mathcal{D})^{-1}\,.
\ee

The analysis of multipoint correlators is straightforward, leading to the general expression
\bea 
&&\langle \Phi^{\dag}(y_1)\dots \Phi^{\dag}(y_l)\Phi(x_l)\dots\Phi(x_1) \rangle =\label{11} \\
&&\frac{1}{\det_{\mathcal{D}}(1+2\mathcal{C}_\mathcal{D})}\sum_{\sigma}\prod_{j=1}^l\left(\frac{\mathcal{C}_\mathcal{D}}{1+2\mathcal{C}_\mathcal{D}}\right)(y_j,x_{\sigma(j)}) \nonumber \, ,
\eea
where the sum runs over all the permutations $\sigma$.

Computing bosonic correlators in the assumption of a gaussian ensemble in the fermionic fields leads to similar expressions, with the introduction of extra minus signs due to the fermionic Wick Theorem
\bea\label{11bis}
&&\langle \psi^{\dag}(y_1)\dots \psi^{\dag}(y_l)\psi(x_l)\dots\psi(x_1) \rangle =\label{eq11} \\
&&\det_{\mathcal{D}}(1-2\mathcal{F}_\mathcal{D})\sum_{\sigma}\text{sign}(\sigma)\prod_{j=1}^l\left(\frac{\mathcal{F}_\mathcal{D}}{1-2\mathcal{F}_\mathcal{D}}\right)(y_j,x_{\sigma(j)}) \nonumber\,,
\eea
where now $\mathcal{F}_\mathcal{D}$ is the fermionic two point correlator restricted to the domain $\mathcal{D}$. Fredholm-like expressions for the bosonic correlators based on a gaussian fermionic ensemble were already known\cite{Lenard,Korepin,kcc14,mckc14}, but with the inverse formula Eq. (\ref{11}) it is now possible to determine the initial conditions of the free (in terms of the fermions) evolution.

\begin{figure*}[t!]
\includegraphics[width=0.33\textwidth]{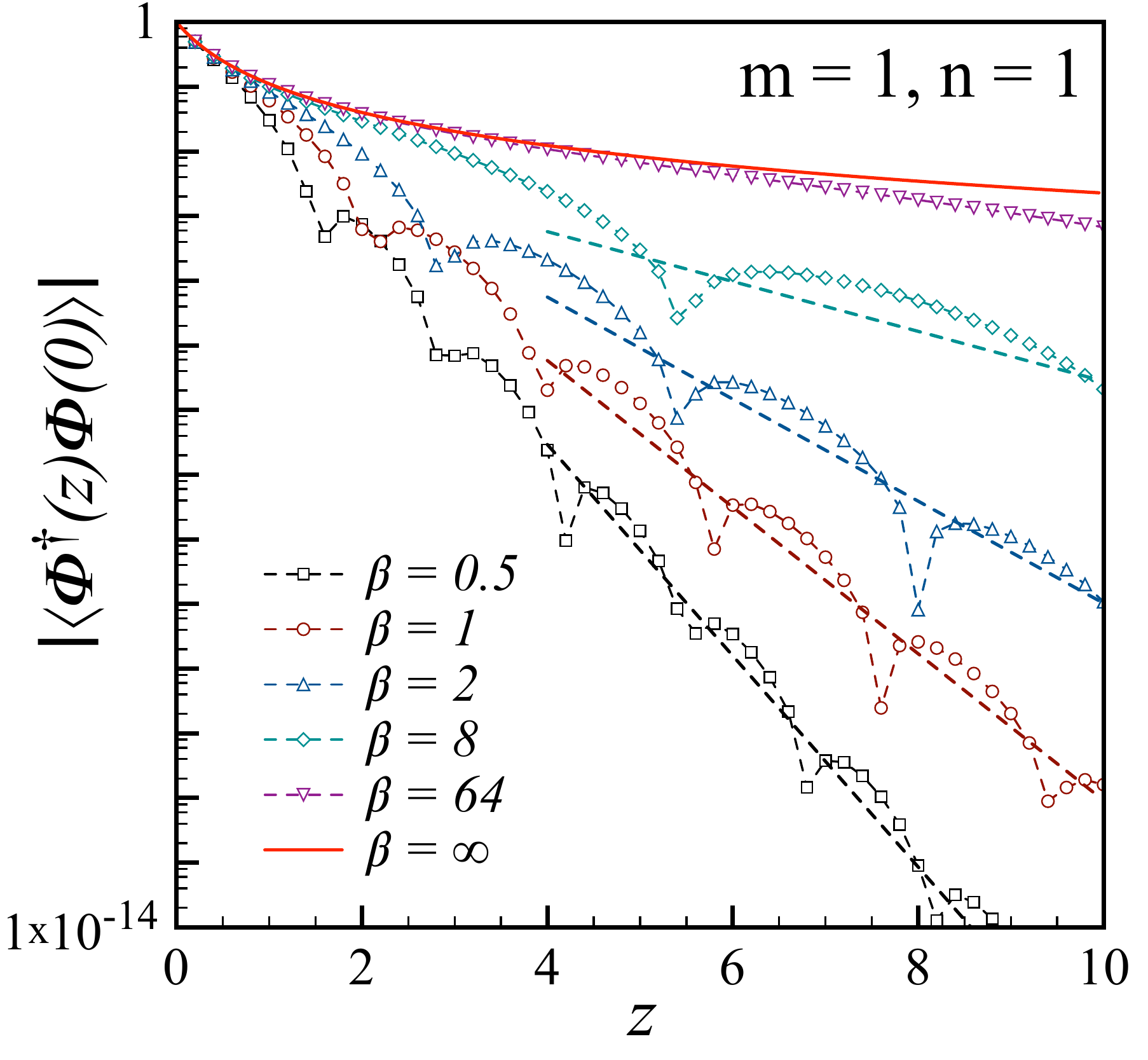}\includegraphics[width=0.33\textwidth]{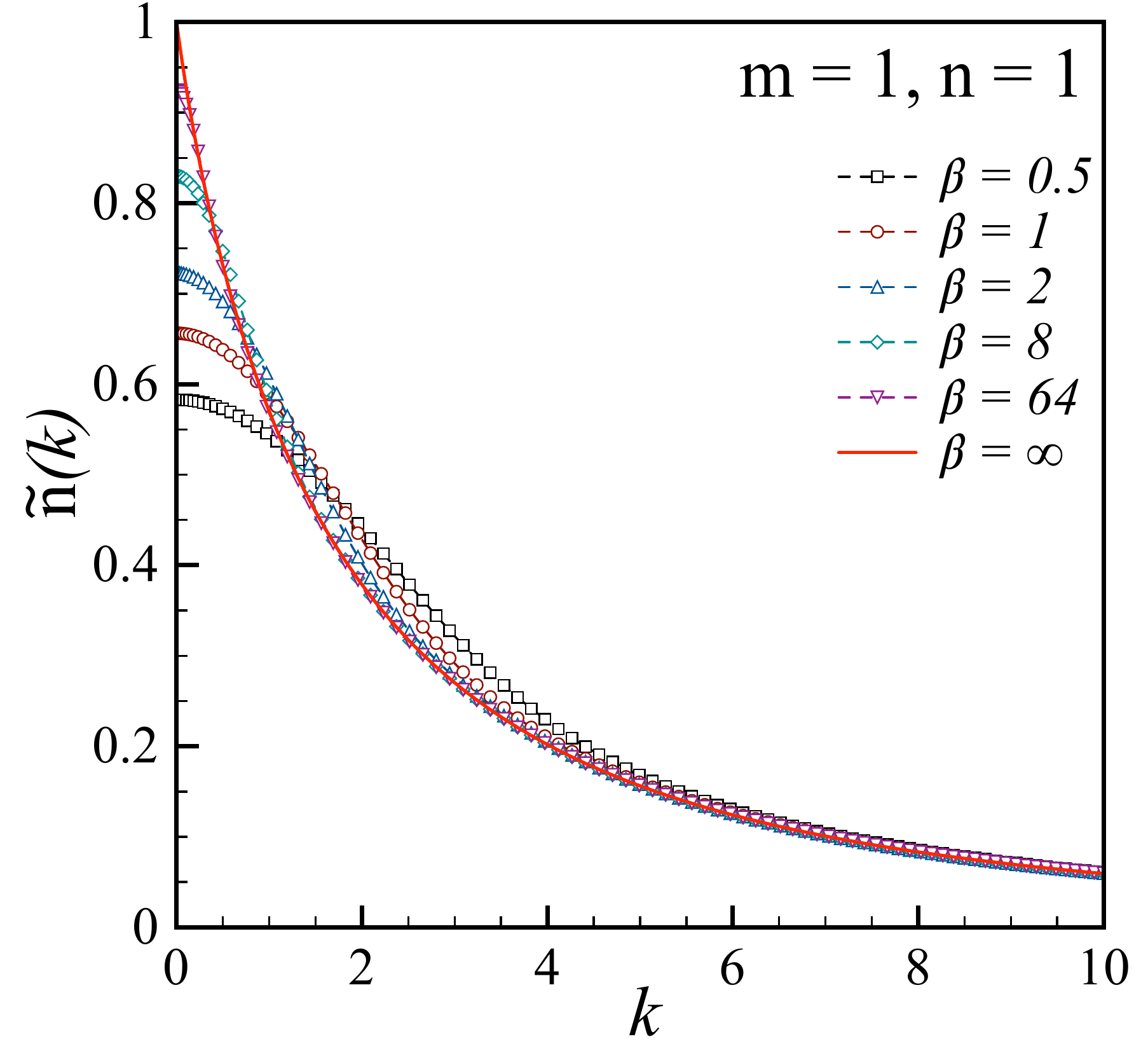}\includegraphics[width=0.33\textwidth]{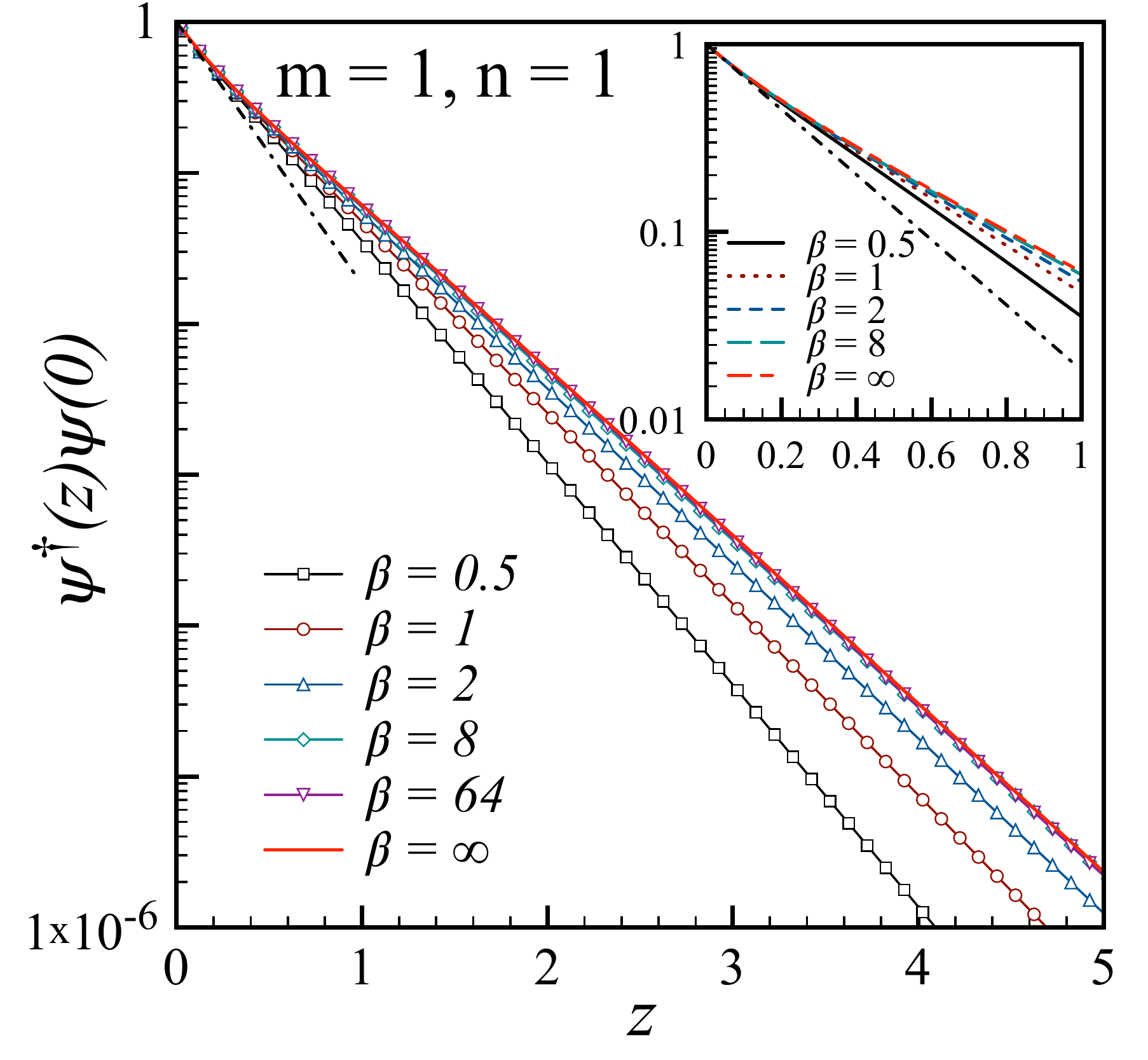}
\caption{\label{fig1}
(Left) Stationary fermionic two-point correlation function
after an interaction quench ($c=0 \, \to \, c=\infty$) 
of a Bose gas initially prepared in a thermal state with inverse temperature $\beta$.
Symbols are the numerical evaluation of Eq. (\ref{11}) with initial bosonic correlator
given by (\ref{thermal_boson}).
In the limit of zero temperature, the correlation function approaches the 
analytic result (\ref{zero_fermion}) (full red line). However, for any finite temperature,
the large-distance exponential decay is captured by the asymptotic expansion
(\ref{asympt_fermion}) (dashed lines).
(Center) Stationary momentum density distribution of the fermions obtained by Fourier transform 
$\langle \Phi^{\dag}(z)\Phi(0) \rangle$.
(Right) Stationary bosonic two-point correlation function obtained by numerical evaluation
of Eq. (\ref{11bis}) with initial fermionic correlator given by $\langle \Phi^{\dag}(z)\Phi(0) \rangle$.
Notice the large distance exponential decay. Nevertheless, such decay is not exponential for all distances.
Indeed, by comparing the correlator with the exact exponential decay which matches the short distance expansion,
namely $n {\rm e}^{- 4 n^2 |x|}$ (dot-dashed black line), one reveals an appreciable discrepancy (see the inset for 
a zoom on the short distances).
}
\end{figure*}

\section{Homogeneous quenches in the Tonks gas phase}\label{sec4}
As first application of the derived expressions we consider the quench protocol in the homogeneous case towards the Tonks gas. For the seek of concreteness, we often refer to the free thermal ensemble as the pre-quench state
\be\label{thermal_boson}
\mathcal{C}(x,y)=\int \frac{dk}{2\pi}\frac{e^{ik(x-y)}}{e^{\beta\left(\frac{k^2}{2m}-\mu\right)}- 1}\, ,
\ee
where the chemical potential $\mu$ is fixed requiring $\mathcal{C}(x,x)=n$, with $n$ the particle density. The steady state information is completely encoded in the initial two point fermionic correlator, that is conserved through the time evolution. While the numerical evaluation of Eq. (\ref{11}) can be easily implemented, its analytical solution can be obtained only in very special cases, since the presence of the finite domain $\mathcal{D}$ prevents the possibility of diagonalizing the operators in the Fourier space. However, Eq. (\ref{11}) is still feasible of proper analytical approximations in several regimes.

\paragraph{Short distances.---} 
In the short distance regime or in the small density limit, a satisfactory approximation of Eq. (\ref{11}) can be obtained from the series expansion in terms of $\mathcal{C}_\mathcal{D}$. 
For example, assuming low density and that the two point bosonic correlators decays fast enough to ensure the convergence of the integrals in Eq. (\ref{5}-\ref{6}) for infinite domain $\mathcal{D}$, we can readily write
\be
\langle \Phi^\dagger(y)\Phi(x)\rangle
= \mathcal{C}(y,x)e^{-2|x-y| \left(n+\mathcal{O}(n^2)\right) }+\mathcal{O}(n^2)\,\,\, .
\ee

\paragraph{Small temperatures.---}
When the two point correlator is approximately constant $\mathcal{C}(x,y)\simeq \mathcal{C}(x,x)=n$ over the whole domain $\mathcal{D}$. This is an important case, since this condition is met in the zero temperature limit, at fixed density, of the free thermal ensemble: in the low temperature limit $\mathcal{C}(x,y)\simeq n e^{-\alpha |x-y|}$ with $\alpha=m(n\beta)^{-1}$. Thus in the small temperature limit $\alpha$ approaches zero and $\mathcal{C}$ is approximatively constant on an increasing range of distances. In this limit the following two point fermionic correlator is immediately derived
\be\label{zero_fermion}
\langle \Phi^\dagger(y)\Phi(x)\rangle=\frac{n}{(1+2n|x-y|)^2}\, .
\ee
This correlator greatly differs from the findings in Ref. \ocite{kcc14} for the BEC case, where the decaying has been found to be purely exponential rather than algebraic. The two results are not in contradiction, since the condensation of one dimensional systems is realized only exactly at zero temperature, thus the zero temperature limit of the thermal ensemble does not reproduce the BEC state.

\paragraph{Large distances.---}
When the distance of the two fermionic operators in the two point correlator far exceeds the typical decay length of the bosonic correlator $\mathcal{C}$, we can extract the asymptotics of the two point fermionic correlator.
We start considering the string contribution
\be
\det_{\mathcal{D}}(1+2\mathcal{C}_\mathcal{D})^{-1}
=e^{-\int_x^y dz \left[\log(1+2\mathcal{C}_\mathcal{D})\right](z,z) } \, ,
\ee
where we already specialized the domain $\mathcal{D}=(x,y)$. In the limit of a very large interval, most of the contribution to the integral comes from coordinates far from the edges of the domain. Thus, in first approximation, we can compute $[\log(1+2\mathcal{C}_\mathcal{D})](z,z)$ as if the domain $\mathcal{D}$ were the whole real axis and diagonalize the operator in the Fourier space. This approximation gives the extensive part of the integral
\be
\det_{\mathcal{D}}(1+2\mathcal{C}_\mathcal{D})^{-1}
=e^{-|x-y|\int \frac{dk}{2\pi}\log\left[1+2\tilde{\mathcal{C}}(k)\right]
+\mathcal{O}\left(|x-y|^0\right)}\, , 
\ee
where $\tilde{\mathcal{C}}(k)$ is the Fourier transform of the two point bosonic correlator. 
The same approximation can be justified on the whole two point fermionic correlator, 
being reliable only in the asymptotic regime
\bea
\langle \Phi^\dagger(y)\Phi(x)\rangle &\propto&  e^{-|x-y|\int \frac{dk}{2\pi}\log\left[1+2\tilde{\mathcal{C}}(k)\right]} \\
&\times&  \int\frac{dk}{2\pi} e^{ik(x-y)}\frac{\tilde{\mathcal{C}}(k)}{1+2 \tilde{\mathcal{C}}(k)}\, . \nonumber
\eea
In the case of the thermal ensemble, we have
\be\label{asympt_fermion}
\langle \Phi^\dagger(y)\Phi(x)\rangle  \propto 
e^{-\xi |x-y|} \left(\frac{e^{-\lambda |x-y|}}{\lambda}+\text{c.c.}\right)\, ,
\ee
where the parameters $\xi$ and $\lambda$ are
\bea
\xi & = & \int \frac{dk}{2\pi}\log\coth\left[\frac{\beta}{2}\left(\frac{k^2}{2m}-\mu\right)\right]\, , \label{decl} \\
\lambda & = & \sqrt{m}\left(\sqrt{|\mu|+\sqrt{\mu^2+\frac{\pi^2}{\beta^{2}}}}+ i\sqrt{-|\mu|+\sqrt{\mu^2+\frac{\pi^2}{\beta^{2}}}} \right)\, . \nonumber
\eea
The analysis of the asymptotes displays an oscillating behavior due to the imaginary part of $\lambda$, however in view of our crude approximations only the overall exponential decay  $\langle \Phi^\dagger(y)\Phi(x)\rangle  \propto e^{-(\xi+\Re(\lambda))|x-y|}$ is expected to be reliable.
In  Fig. \ref{fig1} we plot the numerical computation of the stationary two point fermionic correlator
after a quench from a thermal bosonic ensemble (\ref{thermal_boson}) and compare 
it with our analytical predictions. As already stated before, we only report the pure exponential decay.
Notwithstanding, we checked the predicted period $2\pi/\Im(\lambda)$
against the numerical results and, quite remarkably, we found a very good agreement. 
However, the asymptotic expansion does not give a correct prediction for the oscillations' phase.

We also analyze the momentum density distribution
$\tilde n(k) = \int dz \, {\rm e}^{-ik z} \langle \Phi^\dagger(z)\Phi(0)\rangle$, which turns out to be
very different from the fermionic momentum distribution at thermal equilibrium.
In general, for any finite temperature, the momentum density distribution 
is an analytic function for all momenta. Only in the zero temperature limit, the algebraic decay
of the fermionic two-point function affects the short-distance behavior of the momentum
distribution which reads $\tilde n(k) \simeq 1 - \pi |k|/4n$ (for $\beta\to\infty$).
On the contrary, the large momentum behavior $\tilde n(k)\simeq 8 n^2 /k^2$ 
it turns out to be independent of the temperature, 
being fixed by the non analyticity of the fermionic correlator $\langle \Phi^\dagger(z)\Phi(0)\rangle \simeq n -4 n^2 |z|$ in $z=0$. Let us point out that
such non-analytic behavior in the fermionic two point function 
is somehow {\it universal} since it only depends on the properties 
of the initial bosonic correlator in the vicinity of $z=0$,
namely $\mathcal{C}(x,x)=n$
and $\partial_{x}\mathcal{C}(x,y)|_{y=x} = -\partial_{y}\mathcal{C}(x,y)|_{y=x} = 0$.

Finally, using the steady-state of the two-point fermionic correlation function, 
we can numerically evaluate the stationary bosonic correlators 
$\langle\psi^{\dag}(z)\psi(0)\rangle$ via Eq. (\ref{11bis}).
The Bose gas in the steady state is fully characterized
by a two-point function which exhibits a nearly exact exponential 
decay for all temperatures, even for $\beta=\infty$, 
even tough the stationary fermionic two-point function
decays algebraically. Actually, the exponential decay of the bosonic correlator in the steady state 
can be argued by mean of the symmetric analysis performed 
to extract the asymptotic decay of the two-point fermionic correlator.
Interestingly, for all temperatures, we can easily extract the short distance behaviour
of the bosonic two-point function which, in particular, coincides with the fermionic one, i.e.
$\langle\psi^{\dag}(z)\psi(0)\rangle \simeq \langle \Phi^\dagger(0)\Phi(0)\rangle + 
\partial_z\langle \Phi^\dagger(z)\Phi(0)\rangle|_0 z = n - 4 n^2 |z|$.

Lastly, we remark that even though the asymptotic behavior of the two point fermionic correlator does not suffice to determine the decay of an arbitrary bosonic correlator because of the presence of the string, this is instead possible when the string is absent. This is the case, for example, in the density-density correlator, being the density equivalently expressed in terms of either the fermions or the bosons.
\be
\langle n(x,t) n(y,t)\rangle_\text{steady state}=n^2-|\langle \Phi^\dagger(x)\Phi(y)\rangle|^2
\ee
Thus, the density-density correlator (connected part) decays twice as fast as the two point fermionic correlator (\ref{asympt_fermion}) and it can be computed exactly in the zero temperature limit (\ref{zero_fermion}).

\begin{figure*}[t!]
\includegraphics[width=0.99\textwidth]{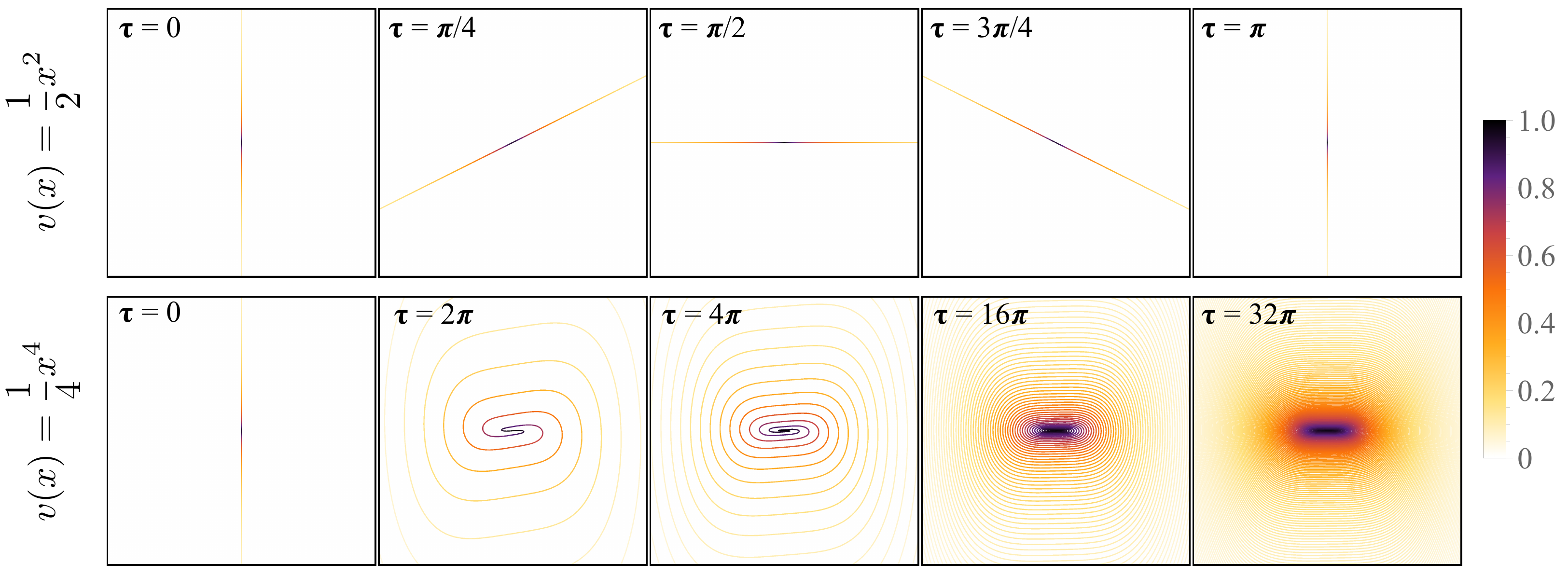}
\caption{\label{fig2}
Color density plot of the Wigner distribution $F_{\tau}(p,q)$ vs $(q,p)\in[-4,4]\times[-8,8]$,
for different times $\tau$ and two representative traps (here $m=1$ and $n=1$). 
The system is initially prepared at zero temperature with $F_{0}(p,q)$ given by Eq. (\ref{F0_zerotemperature}).
In the harmonic case each phase-space trajectory is characterized by the same period 
thus preserving the shape of the initial Wigner distribution whose support simply rotates in time.
In the anharmonic case, the typical dynamical period $T(p_0) \simeq \int_{0}^{q^*} dq/\sqrt{p_0^2 -2mV(q)}$
(with $q^*$ being the de real positive root of $V(q^*) = p_0^2/2m$) depends on the initial energy, and the $\delta$-support of the Wigner distribution
folds over time. For large times, it tends to cover the entire phase plane. 
}
\end{figure*}

\section{Quenching in a thermodynamic trap}\label{sec5}
The expression (\ref{11}) for the correlators does not require translational invariance, but only gaussianity of the initial state. Thus, we can generalize the previous calculations to the more realistic situation in which a confining potential is present and replace the Hamiltonian (\ref{LL}) with
\be
H_{LL}\to H_{LL}+ \int dx \,\,V(x)\psi^\dagger(x)\psi(x),
\ee
where $V(x)$ is a trapping potential. Differently from the homogeneous case, the two point fermionic correlator exhibits now a non-trivial time evolution. 

In presence of the trap, the discrete spectrum prevents the relaxation to a steady state, that can be recovered only in a proper thermodynamic limit: for this reason, we introduce a thermodynamic length-scale $L$. 
Parametrizing the confining potential as $V(x)=v(xL^{-1})$, where $v$ is assumed to be a smooth function kept fixed in the 
thermodynamic limit, the eigenvalues approach a continuum distribution when $L\to\infty$.
While considering the limit of large trap, we must also rescale the number of particles consistently and keep the total density $n=L^{-1}\int dz \,\, \mathcal{C}(z,z)$ constant. 

The thermodynamic limit hugely simplifies the subsequent calculations and permits to consider a generic trap. First of all, in the thermodynamic limit the thermal initial state is described by the Local Density Approximation \cite{LDA} (LDA).
In fact, the two point correlator decays on a typical scale (finite for any non zero temperature) much smaller than the thermodynamic length characterizing the variation of $V$. It is convenient to change variables in the two point bosonic correlator $\mathcal{C}(y,x)\to C(s,q)$, defining $s=y-x$ and $q=(x+y)/2L$.
In the LDA we have
\be
C(s,q)=\int \frac{dk}{2\pi}\frac{e^{iks}}{e^{\beta\left(\frac{k^2}{2m}+v(q)-\mu\right)}-1}\,\,\, .
\ee
As is clear from the general expression (\ref{11}), the LDA on the bosonic correlators implies LDA for the fermionic correlators as well, i.e. $\mathcal{F}(x,y)\to F(s,q)$. 

Following the time evolution after the quench, we can clearly distinguish two time scales.

1) On a non-thermodynamic time scale, the system behaves as if it was locally homogeneous, therefore it exhibits gaussification of the correlators on the same time scale as the homogeneous system. In this first phase, the two point correlator does not evolve: even though the ensemble is gaussian, the steady state has not been reached yet.

2) Relaxation can be realized only on a time scale sufficient for the system to explore the whole trap, thus times $t\propto L$ must be considered.
The two point correlator evolves under the Schr\"odinger equation, that in terms of the rescaled variables reads
\be
i\partial_t F_t=\left\{\frac{\partial_s\partial_q}{Lm}-\left[v\left(q+\frac{s}{2L}\right)-v\left(q-\frac{s}{2L}\right)\right]\right\}F_t\, .\label{v523}
\ee

For any finite time $t$, in the limit $L\to\infty$ the two point correlator does not evolve. However, rewriting the differential equation in terms of the rescaled time $\tau=t/L$ and Taylor expanding the potential $v$, we get a non trivial time evolution
\be
i\partial_\tau F_\tau=\left[\frac{\partial_s\partial_q}{m}-sv'(q)\right]F_\tau\, \label{27}
\ee
where we assumed $\tau\ll L$ (i.e. $t\ll L^2$). The above equation is associated with a classical evolution for the two point correlator. This is unveiled in terms of the well known Wigner distribution \cite{wigner}, that amounts to a partial Fourier transform of the correlator
\be
\tilde{F}_\tau(p,q)=\int ds\,  e^{-ips} F_\tau(s,q)\, \label{eq21}.
\ee
In this new notation, Eq. (\ref{27}) can be written in the form of a classical Liouville equation in terms of $q$ and its conjugated momentum $p$
\be
\partial_\tau\tilde{F}_\tau+\{H_\text{cl},\tilde{F}_\tau\}_{p,q}=0\, ,
\ee
where $\{\, ,\}_{p,q}$ are the classical Poisson brackets and $H_\text{cl}$ is 
the classical Hamiltonian $H_\text{cl}=p^2/2m+v(q)$. In fact, the thermodynamic limit we are considering is equivalent to a semiclassical limit on the time evolution of the Wigner distribution \cite{wigner}.

If the frequency of the solution of the classical equation of motion does not have a trivial dependence on the energy, the correlator in the coordinate space reaches a steady state thanks to the dephasing between the different energy shells. The steady state is of course described in terms of the energy density of the initial Wigner distribution
\be\label{F_eq}
\lim_{\tau\to\infty}F_\tau(s,q)=\int \frac{dp}{2\pi}e^{ips}\epsilon\left(\frac{p^2}{2m}+v(q)\right)\, ,
\ee
where the energy density is defined as
\be\label{E_eq}
\epsilon(E)=\frac{\int dpdq\, \delta\left(\frac{p^2}{2m}+v(q)-E\right) \tilde{F}_{0}(p,q)}{\int dpdq\, \delta\left(\frac{p^2}{2m}+v(q)-E\right) }\,\, .
\ee
Based on the integrability of the free fermionic model, the steady state is expected to be described by a GGE constructed out of the modes of the trap: in Appendix \ref{WKBsec} we show that the quantum mechanical prediction coincides with the above semiclassical result, as it should be.
\begin{figure*}[t!]
\includegraphics[width=0.33\textwidth]{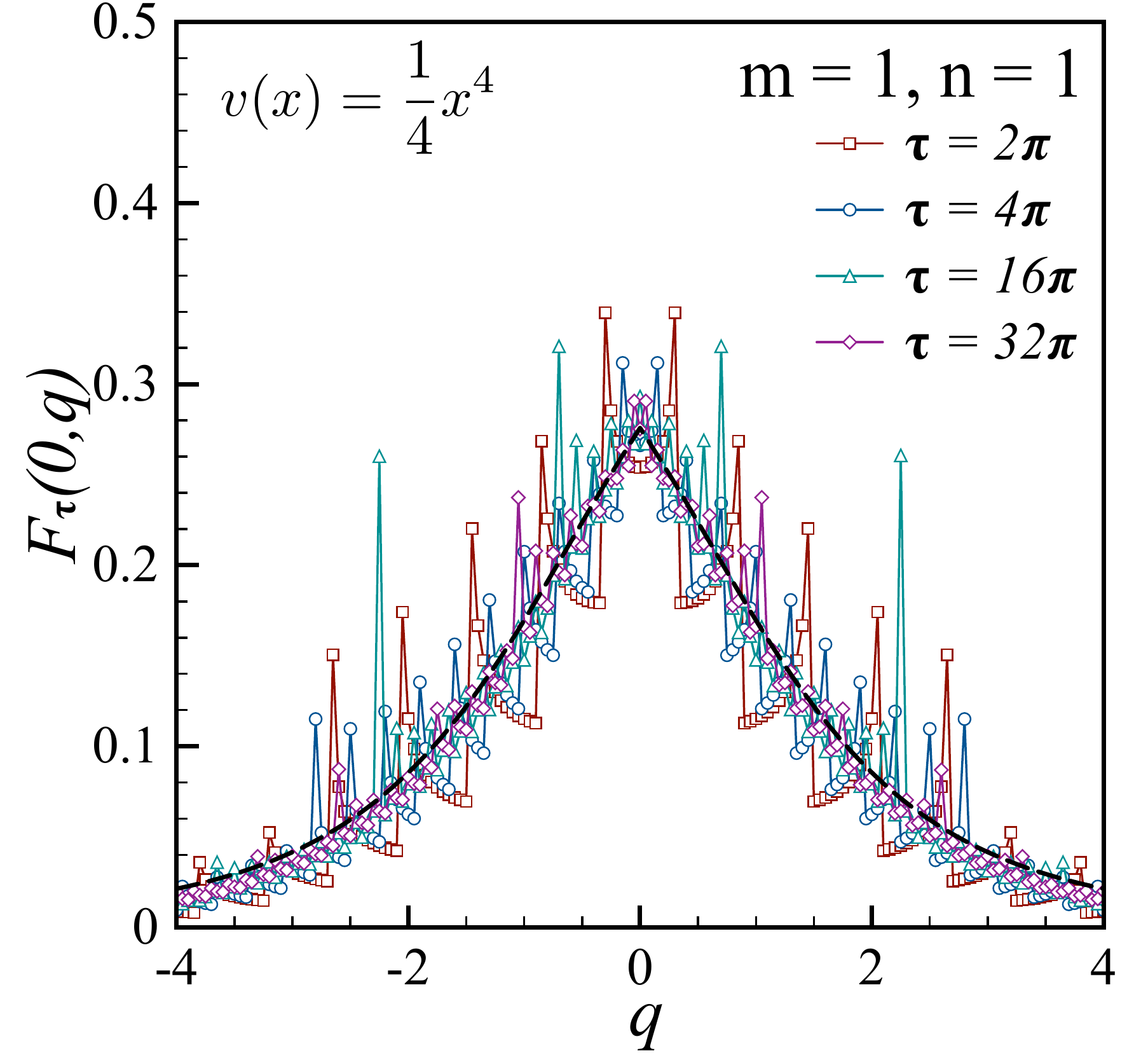}\includegraphics[width=0.33\textwidth]{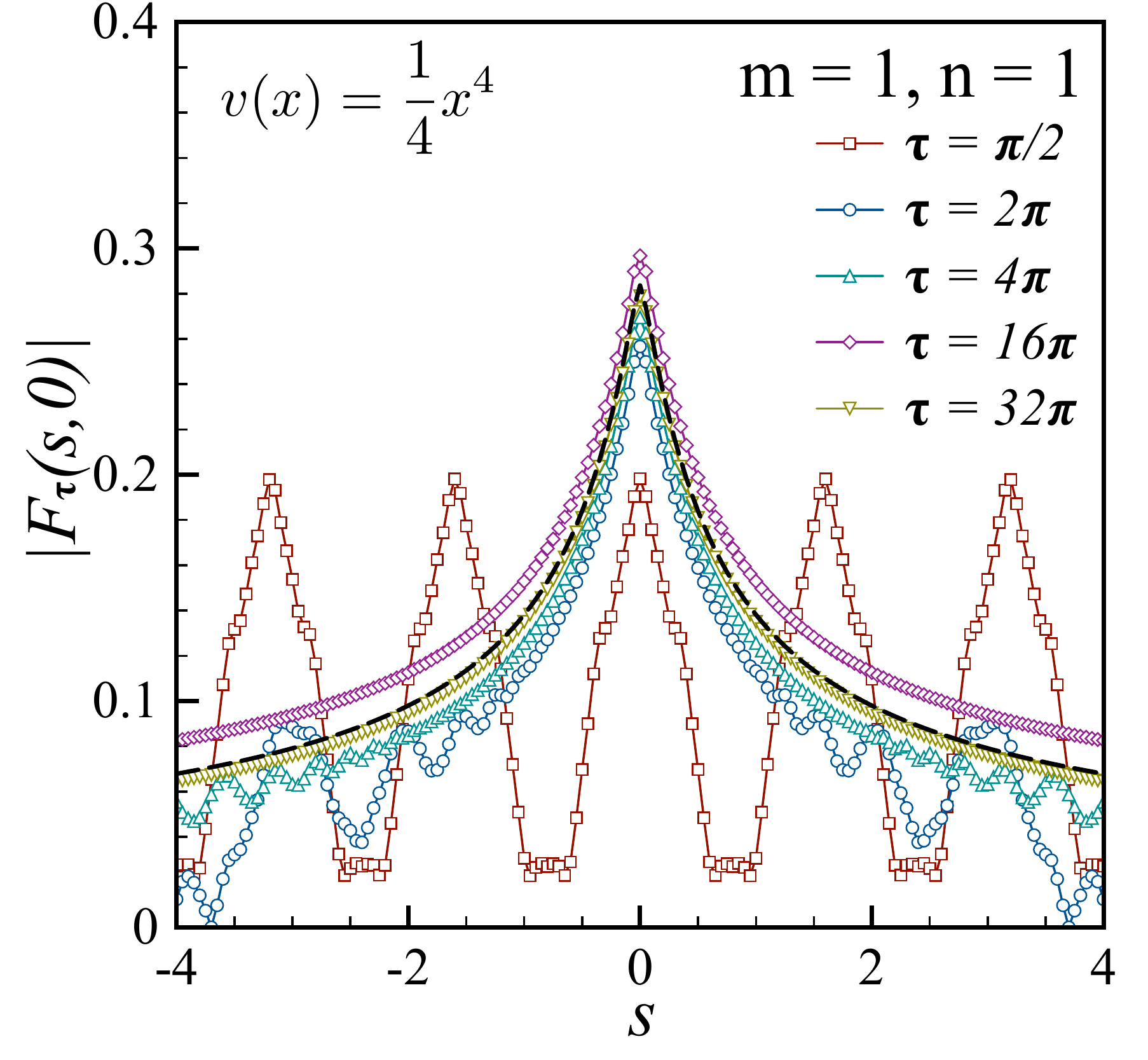}\includegraphics[width=0.33\textwidth]{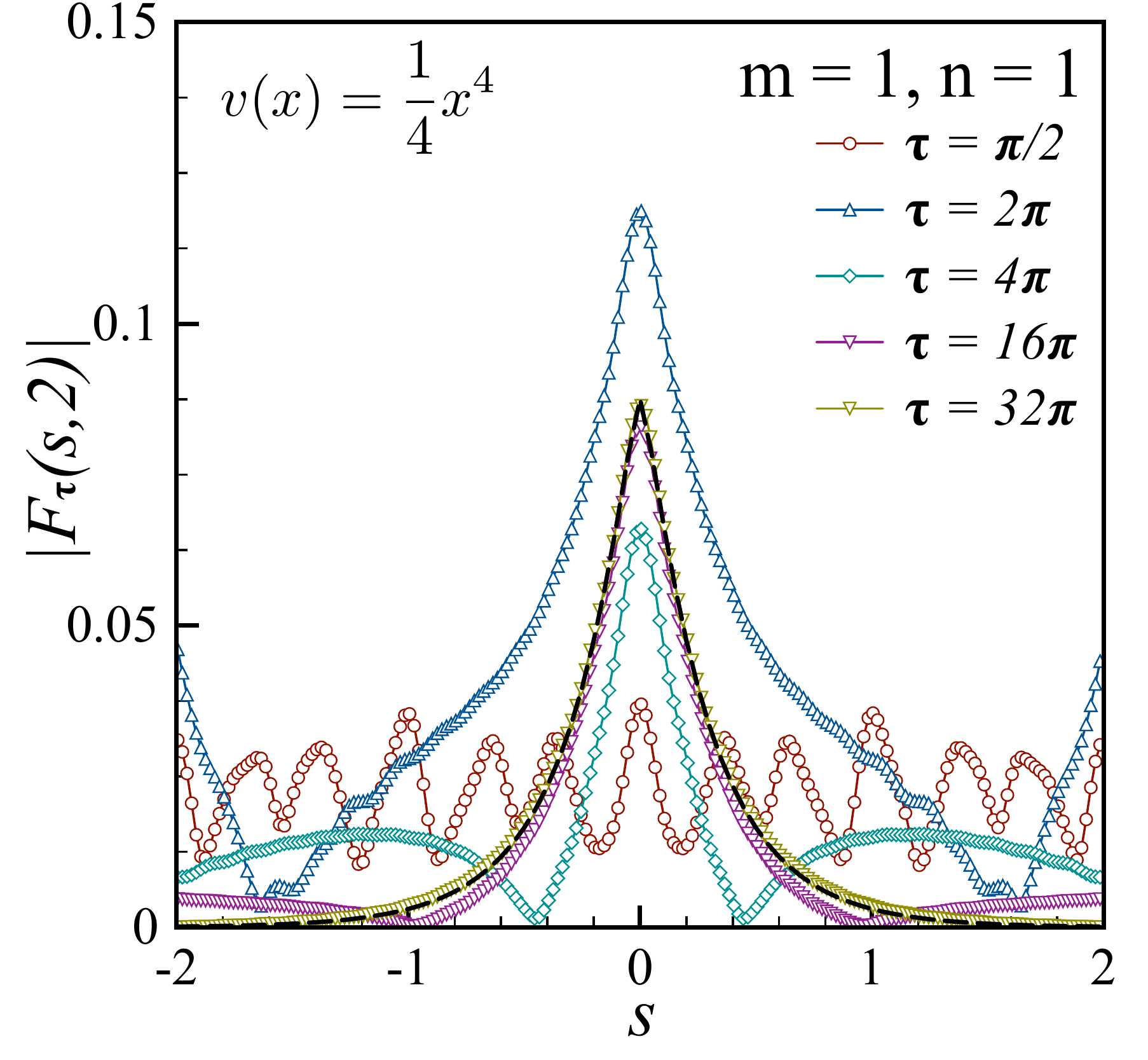}
\caption{\label{fig3}
Zero temperature two-point correlation function for different 
times after an interaction quench in the anharmonic trap. 
In the large time limit the correlator approaches
the steady-state prediction of Eq. (\ref{F_eq}) (black dashed lines).
}
\end{figure*}
As already mentioned, relaxation is caused by the dephasing between the different classical trajectories. While this is the common scenario for a generic trap, this is no longer true in the parabolic case: being the period of the oscillations independent from the initial conditions, dephasing is absent and the two point correlator exhibits revivals. Amusingly, notice that for a parabolic potential Eq. (\ref{v523}) coincides with Eq. (\ref{27}) without taking the thermodynamic limit, thus the Wigner distribution evolves classically even for finite harmonic traps.

As an example, we explicitly analyze the zero temperature limit.
In this limit the LDA of the initial fermionic two-point correlation function reduces to
\be\label{F0_zerotemperature}
F_{0}(s,q) = \delta(q) \frac{n}{(1+2n |s|)^{2}} \, ,
\ee
and the Wigner distribution evolves as
\be
\tilde F_{\tau}(p,q) = \int dp_0 \, \delta(q-q_{\tau}(p_0))\delta(p-p_{\tau}(p_0)) \tilde f (p_0) \, ,
\ee
where $\tilde f(p_0) = \int ds \, {\rm e}^{-ip_0 s} n/(1+2n|s|)^2$, and $(q_{\tau}(p_0),\,p_{\tau}(p_0))$
is the solution of the classical equation of motion with initial condition $(0,p_0)$.
Therefore, exploiting the normalization of the Dirac delta function, we can  
integrate over the final momentum $p$, obtaining for the correlation function
\bea\label{F_sq_num}
F_{\tau}(s,q) & = & \int \frac{dp_0}{2\pi} \, \delta(q-q_{\tau}(p_0)) {\rm e}^{i s p_{\tau}(p_0)} \tilde f (p_0) \\
 & =&   \sum_{j} \frac{1}{2\pi}\left.\frac{ {\rm e}^{i s p_{\tau}(p_0)} \tilde f (p_0)}{|\partial_{p_0} q_{\tau}(p_0)|}\right|_{p_0=p_j(q,\tau)} \, ,\nonumber
\eea
where the sum runs over all the real roots $p_{j}(q,\tau)$ of the equation $q_{\tau}(p_j)=q$. 
In Fig. \ref{fig2} we plot the Wigner distribution at different times $\tau$, 
and for two representative confining potentials, namely $v(x) = x^2 /2$
and $v(x)=x^4 /4$ (where, in the following, we fixed $m=1$ and $n=1$). 
In the harmonic case, the Wigner distribution exhibits a trivial periodic dynamics
and the particle density never reaches a stationary profile, continuously
oscillating between $\delta(q)$ and $\tilde f(q)/2\pi$.

On the contrary, in the anharmonic case,
the number of roots of the equation $q_{\tau}(p_j)=q$ 
increases with time and the root density approaches a continuous distribution;
the system therefore relaxes to the stationary state whose correlation function
is given by Eq. (\ref{F_eq}).

In Fig \ref{fig3} we compare the steady-state predictions with the correlation function evaluated numerically using Eq. (\ref{F_sq_num}), 
in the presence of the anharmonic confining potential $v(x)=x^4/4$.
In particular, we plot the particle density versus $q$ as well as
the two-point correlator as a function of $s = y-x$ 
for fixed $q=0$ and $q=2$. As expected, for large time the 
numerical data approach the stationary profiles. 

\section{Conclusion}\label{conclusion}
In this work we studied the equilibration mechanism of 1D Bose gases in the Tonks-Girardeau limit starting with initial states that are gaussian in the bosonic fields, in the homogeneous case as well as the general inhomogeneous case of an arbitrary confining trap. Despite the noninteracting (in terms of the fermions) dynamics, this problem is nontrivial due to the nonlocal nature of the Jordan-Wigner transformation which links the bosonic and fermionic fields. By solving the problem of extracting fermionic correlations from the bosonic ones in a Gaussian state and vice versa, we derived exact closed formulas for the steady state correlations and analysed them both numerically and analytically in several asymptotic limits.

When a confining potential is considered, the relaxation of local observables can be attained only in a proper thermodynamic limit, where the local density approximation is applicable. In this limit, the evolution is separated on two time scales: initially the system behaves as if it were homogeneous with a subsequent gaussianification. Thereafter, the particles explore the whole trap on a thermodynamic time scale with an emergent semiclassical evolution for the two point fermionic correlator, in terms of which all the expectation values of local observables are determined.

\section{Acknowledgments}
This work was supported by the European Union's Horizon 2020 research and innovation program 
under the Marie Sklodowska-Curie Grant Agreement No. 701221 (M.C.). SS acknowledges support from the A*MIDEX project Hypathie (no. ANR-11-IDEX- 0001-02) and from the LIA - LYSM (Laboratoire Ypatia des Sciences Math\'ematiques).

\appendix

\section{Correlators of fermions on the initial state}
\label{HBsec}
The key point in the analysis of our quench protocol is the mapping between bosonic and fermionic correlators, achieved by mean of the Jordan Wigner transformation (\ref{eq_multipoint_mapping}). This mapping passes through the hard core bosons, whose rigorous treatment requires a lattice regularization. This appendix is devoted to this issue, in particular we show that even though fermionic fields are defined in terms of the hard core bosons, their correlation functions can be computed as if the hard core bosons were standard bosonic fields (after the expressions have been normal ordered). 
In the case when the initial state is chosen to be the Bose-Einstein condensate the explicit check has already been performed in Ref. \ocite{kcc14}, however the same conclusions hold even for more general states. 
Through a simple generalization of the following, it is possible to show the inverse relation, i.e. that bosonic correlators are expressed in terms of fermionic correlators exchanging the role of bosons and fermions in Eq. (\ref{eq_multipoint_mapping}).

Here we test expectation values on pure states with a fixed number of particles, but the generalization over more general density matrices (as the thermal ensemble) is trivial. 
Consider then a generic state $|S\rangle$ with $n$ particles 
\be
|S\rangle=\int d^n x \, W_n(x_1,...,x_n)\psi^\dagger(x_1)...\psi^\dagger(x_n)|0\rangle\, ,
\ee
where $W_{n}$ is the symmetric (normalizable) wavefunction, that we will assume to be smooth. We introduce a lattice space $\Delta$ and discretize the space $x\to \Delta j$, the continuous bosonic fields are replaced with the lattice counterpart $\psi(\Delta j)\to \Delta^{-1/2}\psi_j$
\be
|S_{\Delta}\rangle=\Delta^{n/2}\sum_{j_1,...,j_n} \, W_n(\Delta j_1,...,\Delta j_n)\psi^\dagger_{j_1}...\psi^\dagger_{j_n}|0\rangle\, ,
\ee
where the lattice regularized fields follow standard commutation rules $[\psi_j,\psi_{j'}^\dagger]= \delta_{j,j'}$. To be precise, the lattice wavefunction differs from the continuous one for terms $\mathcal{O}(\Delta)$, but we drop them since they are inessential in the limit in which we are interested in.
The hard core bosons are obtained from the standard bosons by mean of a projector that excludes the presence of more than one particle at each lattice site
\be
\Psi_j=P_j \psi_j P_j\, ,
\ee
where $P_j=|0_j\rangle \langle 0_j|+|1_j\rangle \langle 1_j|$. By mean of a straightforward application of this definition, the hardcore bosonic fields are immediately shown to satisfy the following commutation rules
\be
\{\Psi_j,\Psi^\dagger_{j}\}=0,\hspace{2pc}\Psi_j^2=0 
\ee
\be
[\Psi_j,\Psi_{j'}]=[\Psi_j,\Psi^\dagger_{j'}]=0,\hspace{2pc} j\ne j'
\ee

With the hard core bosons we define the fermions on the lattice through the Jordan Wigner transformation
\be
\Phi_j=\exp\left(i\pi \sum_{j'<j}\Psi^\dagger_{j'}\Psi_{j'}\right)\Psi_{j'}\label{A4}
\ee
and finally, in the $\Delta\to 0$ limit, the continuous fermionic fields $\Phi_j\to \Delta^{1/2}\,\Phi(\Delta j)$. 
Consider now an arbitrary normal ordered correlator of the hard core bosons
\begin{widetext}
\be
\langle S_{\Delta}|\prod_{a=1}^m\Psi^\dagger_{i_a}\prod_{a=1}^m\Psi_{i'_a}|S_{\Delta}\rangle
=\frac{\Delta^{n}(n!)^2}{(n-m)!}\sum_{j_{1},...,j_{n-m}}W^*(\Delta i_1,...,\Delta i_m, \Delta j_1,...,\Delta j_{n-m})W(\Delta i'_1,...,\Delta i'_m, \Delta j_1,...,\Delta j_{n-m})\, ,
\ee
\end{widetext}
where, because of normal ordering, there are not two creation (annihilation) operators with the same index. Moreover, from the sum over the indexes $j$ must be excluded all the configurations where an index $j$ is equal to an index $i$ or $i'$. However, taking the continuous limit, this exclusion requirement can be neglected, since the forbidden region has a zero measure in the integral and we readily obtain
\begin{eqnarray}
&&\label{A6}\langle S_{\Delta}|\prod_{a=1}^m\Psi^\dagger_{i_a}\prod_{a=1}^m\Psi_{i'_a}|S_{\Delta}\rangle=\\
\nonumber&&\Delta^m \left[\langle S|\prod_{a=1}^m\psi^\dagger(\Delta i_a)\prod_{a=1}^m\psi(\Delta i'_a)|S\rangle+\mathcal{O}(\Delta)\right]\, .
\end{eqnarray}

Consider now the fermionic correlators. For the seek of simplicity we analyze the two point case (the extension to the general case is obvious) 
\be\langle S_{\Delta}|\Phi^\dagger_j\Phi_{j'}|S_{\Delta}\rangle=\langle S_{\Delta}|\Psi^\dagger_j:e^{-2\sum_{l=j'+1}^{j-1}\Psi^\dagger_l\Psi_l}:\Psi_{j'}|S_{\Delta}\rangle\, , \label{supphard8}
\ee
where without loss of generality we assumed $j>j'$. Above we rely on a useful identity that permits to normal order the string (\ref{A4})
\be
e^{\sum_l g_l \Psi^\dagger_l\Psi_l}=:e^{\sum_l (e^{g_l}-1)\Psi^\dagger_l\Psi_l}:\,\, .
\ee

We now take the continuous limit of Eq. (\ref{supphard8}), letting $\Delta\to 0$ but keeping $\Delta j=x$ and $\Delta j'=y$. Expanding the normal ordered exponential we get
\begin{eqnarray}
\nonumber&&\langle S_{\Delta}|\Psi^\dagger_j:e^{-2\sum_{l=j'+1}^{j-1}\Psi^\dagger_l\Psi_l}:\Psi_{j'}|S_{\Delta}\rangle=\sum_{n=0}^\infty \frac{(-2)^n}{n!}\\
\nonumber&&\sum_{l_1=j'+1}^{l_1=j'-j}...\sum_{l_n=j'+1}^{l_n=j'-j}\langle S_{\Delta}|\Psi^\dagger_j\Psi^\dagger_{l_1}...\Psi^\dagger_{l_n}\Psi_{l_n}...\Psi_{l_1}\Psi_{j'}|S_{\Delta}\rangle\, .\\
\end{eqnarray}
Aiming to the continuum limit we use (\ref{A6}) in each term and replace the sums with the proper integrals. The resulting expression is then simply resummed leading to
\begin{eqnarray}
&&\langle S_{\Delta}|\Phi^\dagger_j\Phi_{j'}|S_{\Delta}\rangle=\\
\nonumber&&\Delta\left(\langle S|\psi^\dagger(x):e^{-2\int_y^x d\tau \psi^\dagger(\tau)\psi(\tau)}:\psi(y)|S\rangle+\mathcal{O}(\Delta)\right)\, .
\end{eqnarray}
The final step amounts to replace the l.h.s. with the continuous counterpart of the fermionic fields, leading to the desired identity
\be
\langle S|\Phi^\dagger(x)\Phi(y)|S\rangle=\langle S|\psi^\dagger(x):e^{-2\int_y^x d\tau \psi^\dagger(\tau)\psi(\tau)}:\psi(y)|S\rangle\, .
\ee

\section{Quenching in the thermodynamic trap}
\label{WKBsec}
In Section \ref{sec5} we considered the quench protocol in a thermodynamic trap and we showed the emergence of a classical picture for the Wigner distribution, with the subsequent classical relaxation to a steady state.
On the other hand, based on the integrability of the model, we expect relaxation to a GGE constructed out of the modes of the trap. This appendix is devoted to showing the equivalence of the two descriptions, i.e. that the GGE predictions reduce, in the thermodynamical limit, to the classical steady state (\ref{F_eq}-\ref{E_eq}). This consistency check can be achieved thanks to the fact that the WKB approximation \cite{griffiths} becomes exact in the TDL limit we are investigating.

We start presenting a brief review of the WKB method sufficient for our purposes, then use it to compute the GGE predictions in the thermodynamic limit.

\subsection{WKB description of the thermodynamic trap}

Consider the one particle Shroedinger equation in one dimension:
\be
-\frac{\partial_x^2 \varphi(x)}{2m}+V(x) \varphi(x)=E\varphi(x)\, .
\ee
The trapping potential is assumed in the form of Section \ref{sec5}, i.e. $V(x)=v(x L^{-1})$. For simplicity, we assume $v$ to be symmetric and with an unique minimum in $x=0$. In the limit of small variation of the potential the wavefunction can be approximated through the WKB solution. Let $\xi_n>0$ be the classical turning point identified by the equation $E_n=v(\xi_n)$ ($E_n$ is the energy of the $n^\text{th}$ energy level), then the wavefunction oscillates inside the trap ($|x|<L\xi_n$)
\be
\varphi_n(x)= \frac{\mathcal{N}_n \cos\left(\frac{\pi n}{2}+\int_{0}^x dy \sqrt{2m(E_n-v(y/L))}\right)}{[2m(E_n-v(x/L))]^{1/4}}\, \label{supp2}
\ee
and is exponentially damped outside of it. For example, on the right hand side $x>L\xi_n$ we have
\be
\varphi_n(x)\propto\frac{1}{[2m(v(x/L)-E_n)]^{1/4}}e^{-\int_{L\xi_n}^x dy \sqrt{2m(v(y/L)-E_n)}}\, .
\ee
In Eq. (\ref{supp2}) $\mathcal{N}_n$ is a normalization constant to ensure the eigenfunction to be normalized to unity. The energy levels satisfy the WKB quantization condition
\be
\int_{-\xi_n}^{\xi_n}dq \,\sqrt{2m(E_n-v(q))}=\frac{\pi}{L} \left(n+\frac{1}{2}\right)\, .\label{supp3}
\ee
As we already said, the WKB approximation is reliable when the variation of the potential is much smaller than the energy difference between $E$ and the potential, more quantitatively this amounts to ask
\be
\left|\frac{\partial_x^2 V}{(E_n-V)^2}\right|\ll 1,\hspace{2pc}\left|\frac{(\partial_x V)^2}{(E_n-V)^3}\right|\ll 1\, .
\ee
Replacing $V$ with the rescaled potential $v$ (and using the rescaled variable $q=xL^{-1}$) we see that the WKB approximation becomes exact in the TDL 
\be
\frac{1}{L^2}\left|\frac{v''(q)}{(E_n-v(q))^2}\right|\ll 1,\hspace{2pc}\frac{1}{L^2}\left|\frac{(v'(q))^2}{(E_n-v(q))^3}\right|\ll 1\, ,
\ee
with the exception of the turning points, i.e. the zeros of the denominators of the above expressions. However, calling $\delta q$ the distance from the turning point, the region of validity of the WKB approximation is readily estimated as
\be
\frac{1}{|v'(\xi_n)|}\frac{1}{L^2}\ll |\delta q|^{3}\, .
\ee
Thus, in the rescaled variable, the region in which the WKB approximation fails is shrunk to zero as $\delta q\propto L^{-2/3}$. Some care must be used stating that the WKB  becomes exact in the thermodynamic limit, since in terms of the original variable $ x=L q$ the region where WKB fails does not vanish, but rather diverges as $L^{1/3}$. Indeed, even though the single WKB eigenfunction is not reliable in a neighborhood of the turning point of increasing size, the non extensive scaling of the latter permits, ipso facto, a direct use of the WKB eigenfunctions in several expressions.

Before passing to analyze the GGE, we need to properly normalize to unity the wavefunction: this task is hugely simplified in the thermodynamic limit, since only the WKB solution inside the trap is needed. To understand this point consider the solution outside of the trap, let's choose the right side for simplicity. Using the rescaled coordinate $q=L^{-1}x$ the wavefunction is exponentially damped as
\be
e^{-L\int_{\xi_n}^{q} dq' \sqrt{2m(v(q')-E)}}\, \, .
\ee
In the limit $L\to\infty$ the above exponential vanishes, meaning that the region where the wavefunction is not zero outside of the trap does not have a thermodynamic extent.
Using this fact (and the exactness of the WKB approximation in the rescaled coordinate), the normalization constant $\mathcal{N}_n$ in (\ref{supp2}) is readily written as
\be
\mathcal{N}_n=\left[\frac{L}{2}\int_{-\xi_n}^{\xi_n}dq \frac{1}{[2m(E_n-v(q))]^{1/2}}\right]^{-\frac{1}{2}}=\frac{2\sqrt{m}}{\sqrt{L\rho(E_n)}}\,,
\ee
where we neglected subextensive contributions to $\mathcal{N}^{-2}_n$, $\rho(E)$ is simply the volume of the energy shell in the classical phase space
\be
\rho(E)=\int d\tau dp\,\, \delta\left(\frac{p^2}{2m}+v(\tau)-E\right)\, .
\ee
\

\subsection{The excitation density in the thermodynamic limit}

Armed with the WKB approximation, we can start constructing the thermodynamic limit of the GGE associated with the trap. As first step, we extract the excitation energy from the two point fermionic correlator in the assumption the latter satisfied the LDA approximation, i.e. $\langle\Phi^\dagger(x)\Phi(y)\rangle=\mathcal{F}(x,y)\to F(s,q)$ where $s=x-y$ and $q=(x+y)/2L$. $F$ does not contain any further dependence on the thermodynamic length $L$ and decays fast enough for large $s$ to justify the forthcoming approximations.

Denoting as $\eta^\dagger_n,\,\eta_n$ the creation/annihilation operators of the modes of the trap, the number of particles in the energy level $\langle\eta^\dagger_n\eta_n\rangle$ can be readily extracted from the two point fermionic correlator and the eigenfunctions
\be
\langle \eta^\dagger_n\eta_n\rangle=\int dx dy \; \varphi_n^*(x) \varphi_n(y)\mathcal{F}(x,y)\,\, .
\ee
At the price of neglecting corrections vanishing in the thermodynamic limit, we can replace in the above the WKB eigenfunctions and the LDA approximation for the correlator. Using (\ref{supp2}) with the truncation of the wavefunction outside of the trap, through simple trigonometric identities we get

\begin{widetext}
\begin{eqnarray}
\langle \eta^\dagger_n\eta_n\rangle &=&\frac{2m}{L\rho(E_n)}\int_{-L\xi_n}^{L\xi_n} dxdy\,\frac{\cos\left(\int_{y}^x dx' \sqrt{2m(E_n-v(x'/L))}\right)}{[2m(E_n-v(x/L))]^{\frac{1}{4}}[2m(E_n-v(y/L))]^{\frac{1}{4}}}F\left(x-y,\frac{x+y}{2L}\right)\\\nonumber
&+&\frac{2m}{L\rho(E_n)}\int_{-L\xi_n}^{L\xi_n} dxdy\,\frac{\cos\left(\pi n+\int_{0}^x dx' \sqrt{2m(E_n-v(x'/L))}+\int_{0}^y dy' \sqrt{2m(E_n-v(y'/L))}\right)}{[2m(E_n-v(x/L))]^{\frac{1}{4}}[2m(E_n-v(y/L))]^{\frac{1}{4}}}F\left(x-y,\frac{x+y}{2L}\right)\, .
\end{eqnarray}
\end{widetext}
Thanks to the LDA approximation for the fermionic correlator and under the assumption that decays in the relative distance $x-y$, the above can be further simplified. In particular, the second term gives contributions suppressed in the thermodynamic limit and the first line can be approximated to
\bea
&&\langle\eta^\dagger_n\eta_n\rangle=\\
&&\int_{-\xi_n}^{\xi_n} dq\int_{-\infty}^\infty ds\,\frac{\cos\left(s \sqrt{2m(E_n-v(q))}\right)}{[2m(E_n-v(q))]^{\frac{1}{2}}}\frac{2mF\left(s,q\right)}{\rho(E_n)}\, . \nonumber
\eea
The above is clearly a smooth function of the energy $\langle\eta^\dagger_n\eta_n\rangle=\epsilon(E_n)$, where
\be
\epsilon(E)=\frac{1}{\rho(E)}\int dp dq\, \delta\left(\frac{p^2}{2m}+v(q)-E\right)\tilde{F}\left(p,q\right)\label{supp14}
\ee
and $\tilde{F}$ is defined in Eq. (\ref{eq21}). Notice that $\epsilon$ coincides with the classical expression (\ref{E_eq}).

\subsection{GGE predictions in the thermodynamic limit}

The GGE density matrix of the free fermion gas is defined over the modes of the trap $\rho_\text{GGE}=e^{-\sum_n\lambda_n \eta^\dagger_n\eta_n}/\mathcal{Z}$, in particular this ensemble is gaussian and it is completely determined by the two point correlator
\be
\langle\Phi^\dagger(x)\Phi(y)\rangle_\text{GGE}=\sum_n \, \varphi_n^*(x)\varphi_n(y)\langle\eta^\dagger_n\eta_n\rangle_\text{GGE}\, ,\label{supp16}
\ee
where the GGE expectation value of the occupancy $\eta^\dagger_n\eta_n$ is equal to the expectation value on the initial state, that reduces to (\ref{supp14}) in the TDL.
Computing (\ref{supp16}) in the thermodynamic limit simply amounts to replace the eigenfunctions with the WKB expressions and the mode occupation with (\ref{supp14}). In the TDL limit the energy levels become dense, thus we would like to replace the sum over the discrete levels with an integral over the energy, being the density of levels at fixed energy extracted from the quantization rule (\ref{supp3}). However, the energy eigenfunctions are not smooth functions of the energy, since they are even (odd) if the quantum number is even (odd), as it is clear from (\ref{supp2}). Therefore, we first split the sum over the even and odd eigenfunctions, then take the continuum approximation of each one. Doing such an operation and recollecting the terms in an unique integral we readily find
\begin{eqnarray}
&&\langle\Phi^\dagger(x)\Phi(y)\rangle_\text{GGE}=\\
\nonumber&&\int \frac{dE}{2\pi}\,2m\frac{\cos\left(\int_{y}^x dx' \sqrt{2m(E-v(x'/L))}\right)}{[2m(E-v(x/L))]^{\frac{1}{4}}[2m(E-v(y/L))]^{\frac{1}{4}}}\epsilon(E)\, .
\end{eqnarray}
Since $\epsilon(E)$ is a smooth function, the above integral decays when $x$ and $y$ are pulled far apart and in the limit of large $L$ we can safely approximate
\be
\langle\Phi^\dagger(x)\Phi(y)\rangle_\text{GGE}=\int \frac{dp}{2\pi} e^{ips}\epsilon\left(\frac{p^2}{2m}+v(q)\right)\, ,
\ee
where, as usual, $s=x-y$ and $q=(x+y)/2L$. The above expression matches with the classical computation (\ref{F_eq}-\ref{E_eq}), as it should be.


\end{document}